\documentclass[aip,reprint,cha]{revtex4-2}
\usepackage{graphicx}
\usepackage{amsmath,amssymb}
\usepackage[usenames]{color}

\usepackage{hyperref}
\hypersetup{colorlinks,
 linkcolor=blue,%
 citecolor=blue,%
 urlcolor=blue
}

\begin{document}

\title{Synchronization transitions in adaptive simplicial complexes with cooperative and competitive dynamics}
 
\author{S. Nirmala Jenifer}
\affiliation{Department of Physics, Bharathidasan University, Tiruchirappalli 620024, Tamil Nadu, India}

\author{Dibakar Ghosh}
\affiliation{Physics and Applied Mathematics Unit, Indian Statistical Institute, 203 B. T. Road, Kolkata - 700108, India}

\author{Paulsamy Muruganandam}
\affiliation{Department of Physics, Bharathidasan University, Tiruchirappalli 620024, Tamil Nadu, India}

\date{\today}

\begin{abstract}
Adaptive network is a powerful presentation to describe different real-world phenomena. However, current models often neglect higher-order interactions (beyond pairwise interactions) and diverse adaptation types (cooperative and competitive) commonly observed in systems like the human brain and social networks. This work addresses this gap by incorporating these factors into a model that explores their impact on collective properties like synchronization. Through simplified network representations, we investigate how the simultaneous presence of cooperative and competitive adaptations influences phase transitions. Our findings reveal a transition from first-order to second-order synchronization as the strength of higher-order interactions increases under competitive adaptation. We also demonstrate the possibility of synchronization even without pairwise interactions, provided there is strong enough higher-order coupling. When only competitive adaptations are present, the system exhibits second-order-like phase transitions and clustering. Conversely, with a combination of cooperative and competitive adaptations, the system undergoes a first-order-like phase transition, characterized by a sharp transition to the synchronized state without reverting to an incoherent state during backward transitions. The specific nature of these second-order-like transitions varies depending on the coupling strengths and mean degrees. With our model, we can control not only when the system synchronizes but also the way the system goes to synchronization. 
\end{abstract} 

\maketitle

\begin{quotation}
Complex networks like brains and social networks can be useful to understand how the interconnected elements within these systems work together. To enhance these models, people explore how groups of elements, rather than just pairs, influence each other. These higher-order interactions can significantly impact how elements synchronize or act together. This study explores how various adaptations, including cooperation and competition, affect synchronization. Intriguingly, our model suggests that synchronization could occur in the presence of strong adaptive higher-order interactions only. Recent studies indicate that explosive synchronization can be achieved even when there is no correlation between the natural frequency and degrees of the oscillators. This study reports on the observation of both explosive transition and continuous transition to synchronization in the presence of adaptions and higher-order interactions. By manipulating how nodes adapt and the coupling strengths, one can get the desired type of transition, such as explosive or continuous. This research could help us develop more accurate models of real-world networks by capturing the complexities of how elements mutually influence each other within groups.
\end{quotation}

\section{Introduction}
\label{sec:1}
Simplicial complexes are one of the representations used to describe complex networks with interactions between more than two nodes or, in other words, higher-order interactions \cite{Battiston2020, Battiston2021, Majhi2022, Iacopini2019}. Complex network theory has long been used to model intricate systems in nature, such as the brain, social networks, and disease outbreaks. These networks are typically visualized as collections of nodes (entities) connected by links (interactions) \cite{Boccaletti2006}. As research has progressed, it has become increasingly recognized that complex networks exhibit not only pairwise interactions (between two nodes) but also, importantly, higher-order interactions involving three, four, or more nodes. The inclusion of these higher-order interactions is crucial for a comprehensive understanding of dynamics within complex systems.

On the other hand, adaptive networks are dynamical systems where links can appear, disappear, or change their weights over time depending on their dynamical states. \cite{ sawicki2023perspectives}. These networks follow various adaptation rules, making them rich in information \cite{berner2023adaptive, kasatkin2017self, berner2019multiclusters}. They provide insights into the evolving interactions within complex systems \cite{berner2021desynchronization}. The collective behaviours that emerge in complex dynamical systems cannot be solely described by the properties of their individual parts, such as nodes and links \cite{liang2021measuring}. Synchronization, for instance, is a notable collective phenomenon studied extensively since its discovery in fireflies. For example, Winfree \cite{winfree1967biological} and Kuramoto \cite{Kuramoto1975, rodrigues2016kuramoto} developed models to analyze synchronization in complex systems, representing networks as a collection of weakly coupled nonlinear oscillators with natural frequencies. The nodes in complex systems adjust their dynamics due to the interaction between the nodes to have a collective common dynamics. This phenomenon is widely known as synchronization \cite{arenas2008synchronization, gomez2007paths}. As the coupling strength increases, the system transitions from an incoherent state to a coherent one, typically in a second-order manner.

Explosive synchronization (ES) is one of the emerging phenomena in complex networks. This phenomenon describes a discontinuous and irreversible transition from an incoherent state to a coherent state, similar to a phase transition in matter. Initially, ES is observed in systems when there is a microscopic positive correlation between the degrees of the nodes and the natural frequencies of the Kuramoto oscillators in scale-free networks \cite{gomez2011explosive}. Subsequently, ES has been observed under various circumstances, such as when coupling strengths are correlated with natural frequencies \cite{zhang2013explosive} and adaptively increased \cite{zhang2015explosive} and in the presence of multiple layers \cite{frolov2021coexistence}. Furthermore, it has been found in complex networks with interdependent and competitive interactions across single-layer and multilayer configurations \cite{frolov2021coexistence} and in. Frolov and Hramov \cite{Frolov2021} demonstrated, in the context of extreme events, that the hierarchical structure of complex networks can cause increased excitability and explosive synchronization. This interplay leads to brief bursts of extreme synchronization, which could be considered distinctive markers of epileptic brain activity. On scale-free networks, ES is also observed with self-organized bistability where the system can switch between the coexisting states \cite{frolov2022self}. Even though much research has been done on adaptive networks, the effects of cooperative and competitive adaptation in higher-order networks are still yet to be studied. 

In this paper, we present a model for systems exhibiting higher-order interactions, including both cooperative and competitive adaptations of nodes participating in pairwise and non-pairwise interactions. We achieve this by leveraging a simplicial complex representation of complex systems. We then analyze how the form of adaptation in links and triangles affects synchronization transitions and provide analytical proof for the occurrence and disappearance of explosive synchronizations. We also discuss the effect of the number of nodes and mean degree in the adaptive networks on the transition of synchronization.

\section{The model}
\label{sec:2}
We consider a system of Kuramoto oscillators with competitively and collaboratively adaptive coupling. Additionally, our model incorporates higher-order interactions, meaning that nodes interact not only through pairwise but also through triangular (three nodes interaction) configurations. The mathematical form of the network equation is represented as,
\begin{align}
\dot \phi_{i} = &\, \omega_i + \alpha_i {\sigma_1}\sum_{j=1}^N a_{ij}^{(1)} \sin(\phi_j - \phi_i ) \notag \\ 
& \, + \beta_i{\sigma_2} \sum_{j=1}^N \sum_{p=1}^N a_{ijp}^{(2)} \sin(2 \phi_j - \phi_i - \phi_p ). \label{eq:1} 
\end{align}
Here, $\omega_i$ is the natural frequency of the $i$-th oscillator, which can be distributed randomly from $[-1,1]$. $\dot\phi_i$ is the instantaneous frequencies of the oscillator. $\sigma_1$ and $\sigma_2$ are the coupling strengths of the pairwise and the three-body interactions, respectively. $a_{ij}$s are the elements of the adjacency matrix $A^{(1)}$. $a_{ji} = 1$, if two nodes $i$ and$ j$ are connected and $a_{ij} = 0$, if $i$ and $j$ are not connected. $a_{ijp}$s are the elements of the adjacency tensor $A^{(2)}$. $a_{ijp} = 1$, when the nodes $i, j$ and $k$ have three-body interactions. If they do not involve in the three-body interaction, $a_{ijp} = 0$.
We consider $f_1$ and $f_2$ as fractions of oscillators, which can be either cooperatively interacting or competitively interacting. We consider $\alpha_i = 1 - r_i$, for $i = 1, 2, \ldots, Nf_1$, i.e., nodes interact via links coupled competitively and the rest number of nodes $i = Nf_1 + 1, Nf_1+2,\ldots, N$, $\alpha_i = r_i$, i.e., they are coupled cooperatively. Similarly, for $i = 1, 2, \ldots, Nf_2$, $\beta_i = 1- r_i$ nodes interact competitively via triangles and $i = Nf_1 + 1, Nf_2+2,\ldots, N$, $\beta_i = r_i$ and nodes in triangles are cooperatively coupled. When $f_1 = 0 $ and $f_2 = 1$, the pairwise links are coupled cooperatively and the nodes that form triangles are coupled competitively. The opposite situation has occurred when $f_1 = 0$ and $f_2 = 1$. Here, $r_i$ is the local order parameter of the $i$-th oscillator and given by,
\begin{align}
r_i \exp{(\mathrm{i} \phi_i)}  = &\, \frac{1}{k_i} {\sum_{j=1}^N a_{ij}^{(1)} \exp{(\mathrm{i} \phi_j)}}, \label{eq:1b}
\end{align}
where $\mathrm{i}=\sqrt{-1}$ and $k_i$ is the degree of the $i$-th node. The cooperative oscillators contribute more to synchronization as the system starts to synchronize. Competitive oscillators prevent synchronization as the system tends to achieve a coherent state. With this model, we can model many real-world networks, such as social networks, the spreading of disease and the brain, where the nodes in the system interact with other nodes, not only via pairwise interactions, but also via non-pairwise interactions. Moreover, the nodes participating in the links and triangles adapt as the system evolves. In those networks, the presence of cooperative oscillators accelerates synchronization as the system starts to synchronize, and competitive oscillators try to prevent synchronization. In natural systems, there will always be the presence of nodes that will contribute cooperatively and competitively to the system synchronization. In some cases, nodes that are competitive in links could be cooperative in non-pairwise interactions and vice versa. Thus, we try to consider different possible combinations of adaptations and how they affect the synchronization transitions.

\section{Numerical Results}
\label{sec:3}
For the numerical analysis, we consider an Erd\"os-R\'enyi (ER) network of $200$ oscillators with $20$  average degree. The nodes in the network adapt as the system synchronizes either cooperatively or competitively or both. Depending on the nature of adaptation, we consider the following four cases:
\begin{enumerate}
\item[(i)] nodes in both pairwise and three-body interactions adapt cooperatively,
\item[(ii)] nodes in links adapt cooperatively and the nodes interacting via triangles adapt competitively,
\item[(iii)] nodes in the links adapt competitively and the nodes in triangles adapt cooperatively,
\item[(iv)] nodes in both links and triangles adapt competitively.
\end{enumerate}

In all these cases, we vary the pairwise coupling strength $\sigma_1$ at fixed value of higher-order interaction strength $\sigma_2$ and investigate the effects of higher-order interactions in each cases. In addition, we fix the natural frequencies of the oscillators randomly in the range $[-1, 1]$, and the phases of the oscillators are also taken randomly in the range $[0, 2\pi]$ and assume all the triangles as three-body interactions. To measure the degree of synchronization, we use the global order parameter $R$ as
\begin{align}
R \mathrm{e}^{\mathrm{i}\psi} & = \frac{1}{N} \sum_{j=1}^{N} \mathrm{e}^{\mathrm{i}\phi_j}. \label{eq:1c} 
\end{align}
The value of $R$ lies in $[0, 1]$. Here, $R=1$ when all the oscillators are synchronized, whereas $R=0$ indicates that all the oscillators are independent and oscillate at their natural frequencies, resulting in an incoherent state. 

\subsection{Pairwise and non-pairwise interactions are cooperative}

\begin{figure}[!ht]
\centering 
\includegraphics[width=0.95\linewidth]{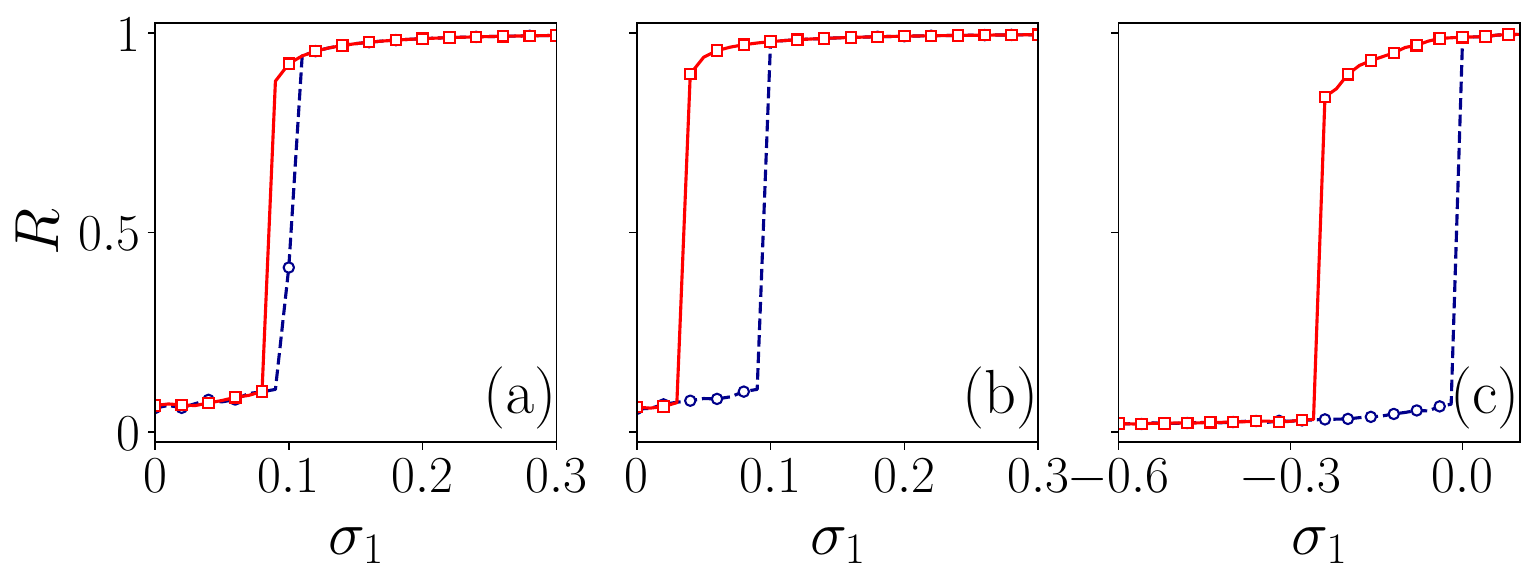}
\caption{Variation of the global order parameter $R$ by changing the pairwise coupling strength  $\sigma_1$ for different values of higher-order interaction strength (a) $\sigma_2 = 0$, (b) $\sigma_2 = 0.05$ and (c) $\sigma_2 = 0.3$. All the links and triangles adapt cooperatively. The blue and red lines represent the forward and backward transition, respectively. As $\sigma_2$ increases, the width of the hysteresis is increasing. When $\sigma_2$ is large enough, the system goes to synchronization even when there are no pairwise interactions, and the system goes back to the incoherent state when the pairwise couplings are repulsive.} 
\label{fig:1}
\end{figure}

To make both interactions cooperatively adaptive, we choose $f_1=f_2= 0.0$. We choose three values of $\sigma_2 = 0.0$, $0.05$, $0.3$ and the values of $R$ as a function of $\sigma_1$ are shown in Figs. \ref{fig:1}(a-c). In the forward transition, we vary the values of $\sigma_1$ with the step of $0.01$ adiabatically from $0.0$. On the backward transitions, we can observe the critical coupling strengths for desynchronization from the coherent state shifts towards the left, and hence the increase in the width of the hysteresis and the system sustains synchronization for a wider range of coupling strengths, as non-pairwise coupling strength increases when the nodes in links and triangles adapt cooperatively. %
When $\sigma_2$ is large enough, the system does not restore to desynchronization in the backward transition even after reaching $\sigma_1 = 0.0$. The above shows that the presence of non-pairwise interactions with nodes adapting cooperatively can maintain synchronization even if there are no pairwise interactions. To get the system back to an incoherent state, we decrease the values of pairwise coupling strengths below zero and make nodes repulsively coupled. The critical value for backward transitions shifted to the negative region (in Fig. \ref{fig:1}(c)), and the hysteresis also widened. When $\sigma_2$ is large enough, it becomes coherent even before zero. The results are shown in Fig. \ref{fig:1}. The number of drifting oscillators reduces, and the number of coherent oscillators that participate in synchronization increases, which may be due to the presence of a higher-order interaction in the expression for $R$. When we increase $\sigma_2$, the turning point shifts towards the lower value of the coupling strengths. 

\subsection{Cooperative pairwise and competitive non-pairwise adaptations}
Next, we consider a case where nodes participate in pairwise interactions adapt cooperatively and the nodes in the triangles adapt competitively. For this purpose, we consider $f_1 = 0$ and $f_2 = 1$. Here, we have found that the transition from an incoherent state to a coherent state happens as a first-order transition with hysteresis for the smaller values of $\sigma_2$. %
\begin{figure}[!ht]
\centering
\includegraphics[width=0.95\linewidth]{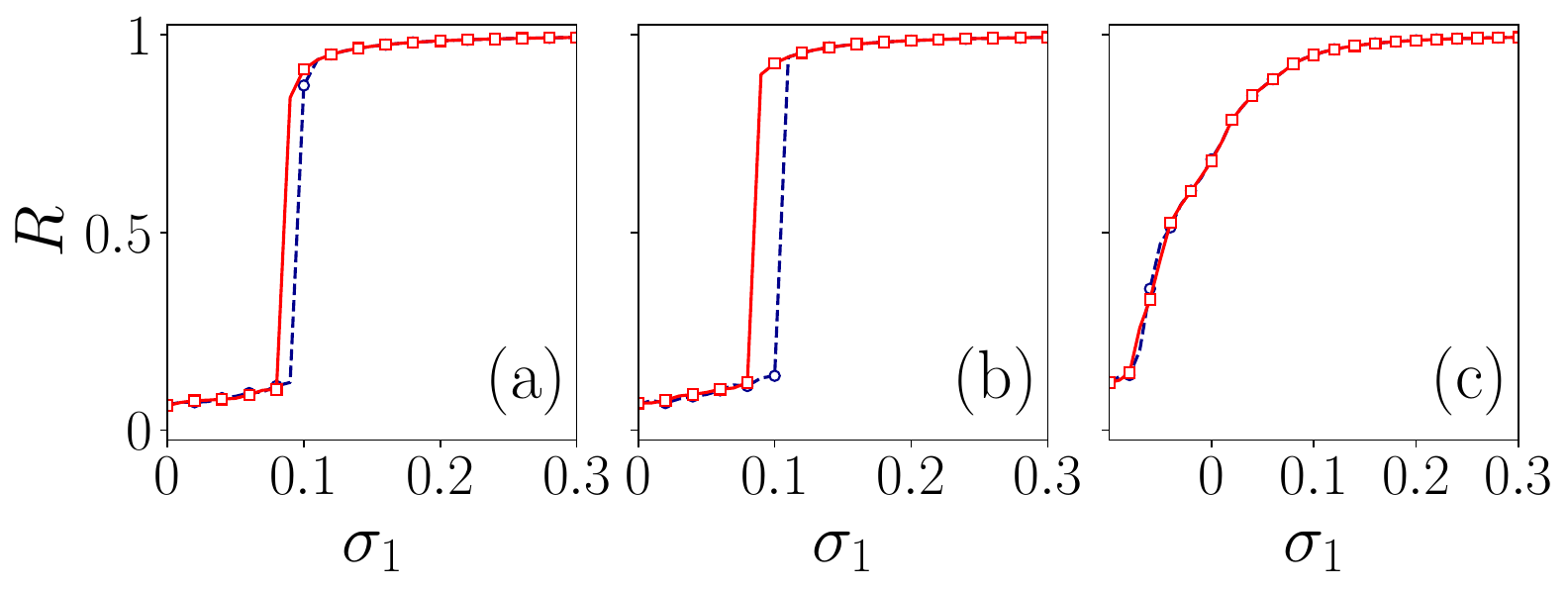}
\caption{Global order parameter  $R$ as a function of $\sigma_1$ calculated for the different values of (a) $\sigma_2 = 0.0$, (b) $\sigma_2 = 0.05$ and (c)  $\sigma_2 =0.3$. All the nodes in links adapt cooperatively, and in triangles adapt competitively. The blue and black lines represent the forward and backward transition, respectively. As $\sigma_2$ increases, the width of the hysteresis increases. When $\sigma_2$ is large enough, the synchronization is not explosive, but rather continuous.} 
\label{fig:2}
\end{figure}
For the higher values of coupling strengths of non-pairwise interactions, the nodes of which adapt competitively, the explosive synchronization vanishes for the same system configuration. The system is partially synchronized even when $\sigma_1 = 0$ and slowly goes to synchronization as $\sigma_1$ increases. For the same value of $\sigma_2 = 0.3$, as in the previous case, when the nodes in non-pairwise interactions adapt cooperatively, the system goes to explosive synchronization, but when the nodes adapt competitively, the system undergoes a second-order transition. The results are shown in Fig.~\ref{fig:2} and as in the above case, as the width of the hysteresis increases, the turning point moves to the lower value of the $\sigma_1$ and the partial synchronization at $\sigma_1 = 0$ when $\sigma_2 = 0.3$, could be observed. 

\subsection{Competitive pairwise and cooperative non-pairwise adaptations}

\begin{figure}[!ht]
\centering
\includegraphics[width=0.95\linewidth]{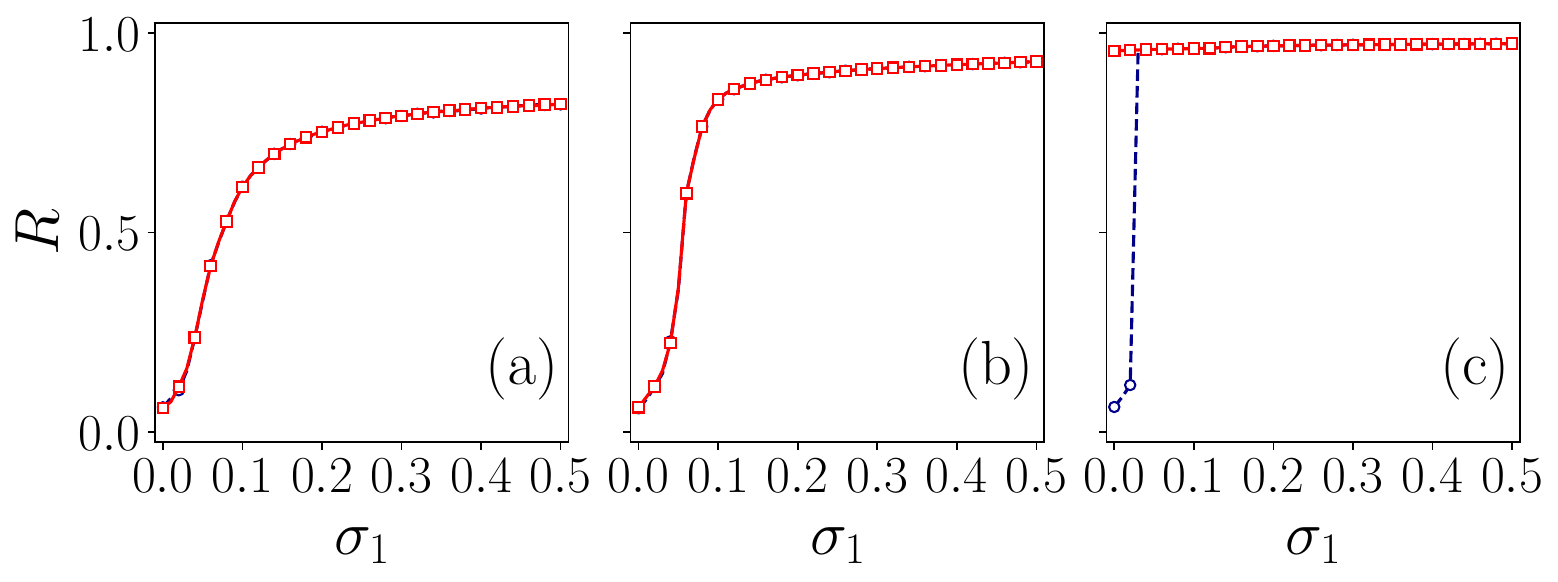}
\caption{Variation of $R$ with respect to $\sigma_1$ for the different values of (a) $\sigma_2 =0.0,$ (b) $\sigma_2 =0.05,$ and (c) $\sigma_2= 0.1$. All the links adapt competitively, and triangles adapt cooperatively. The blue and red lines represent the forward and backward transitions, respectively. As $\sigma_2$ increases, the transition from incoherent to synchrony changes from continuous to discontinuous. When $\sigma_2=0.1$ is large enough, while backward transitions, the system does not come back to desynchrony. The synchronization sustains even when there are no pairwise interactions.}
\label{fig:3}
\end{figure}

To make the node interactions in links competitive and triangles cooperatively adaptive, we choose $f_1= 1.0$ and $f_2=0.0$. In the forward transition, we choose the values of $\sigma_1$ from $0$ to $0.5$ with $\sigma_2$ fixed as $\sigma_2 = 0.0$, $0.05$, and $0.1$ [in Figs.~\ref{fig:3}(a - c)]. In this case, we observe second-order phase transition when there are no higher-order interactions, but the system is only partially synchronized with $R$ approximately equal to $0.7$. %
Initially, the system synchronizes faster because the nodes competitively contribute to the process. However, when the system reaches a partially synchronized state with $R = 0.5$, this competition hinders further progress, slowing the overall synchronization. When we include the higher-order interaction interaction with cooperative adaptivity, the initial rate at which the system goes to synchronization increases and for higher values of $\sigma_2$, the system experiences a jump like transition from an incoherent to a coherent state. But, on the backward transition, the system goes to the incoherent state at the same critical value as in the forward transition. If we increase the $\sigma_2$ more, the system goes to a sharp transition from incoherent to coherent, but on the backward transition, it does not go back to zero value, even in the case of negative coupling of pairwise interactions [see Fig.~\ref{fig:3}(c)]. This behavior is different from the above cases, where the system restores desynchrony when the pairwise interactions are repulsively coupled ($\sigma_1<0$). With $R = 1$, the competitive term vanishes since it adapts as $1 - R$. When the system reaches the synchronized state, only the non-pairwise interaction term contributes to synchronization. This term ceases to adapt when $R$ becomes equal to $1$, and consequently, the system stops adapting. However, the system does not revert to a desynchronized state upon a backward transition because the pairwise interactions vanish after the system achieves synchronization. As a result, synchronization is then solely maintained by the non-pairwise interactions and the result is shown in Fig.~\ref{fig:3}(c).

\subsection{Pairwise and non-pairwise interactions are competitive}
\begin{figure}[!ht]
\centering
\includegraphics[width=0.95\linewidth]{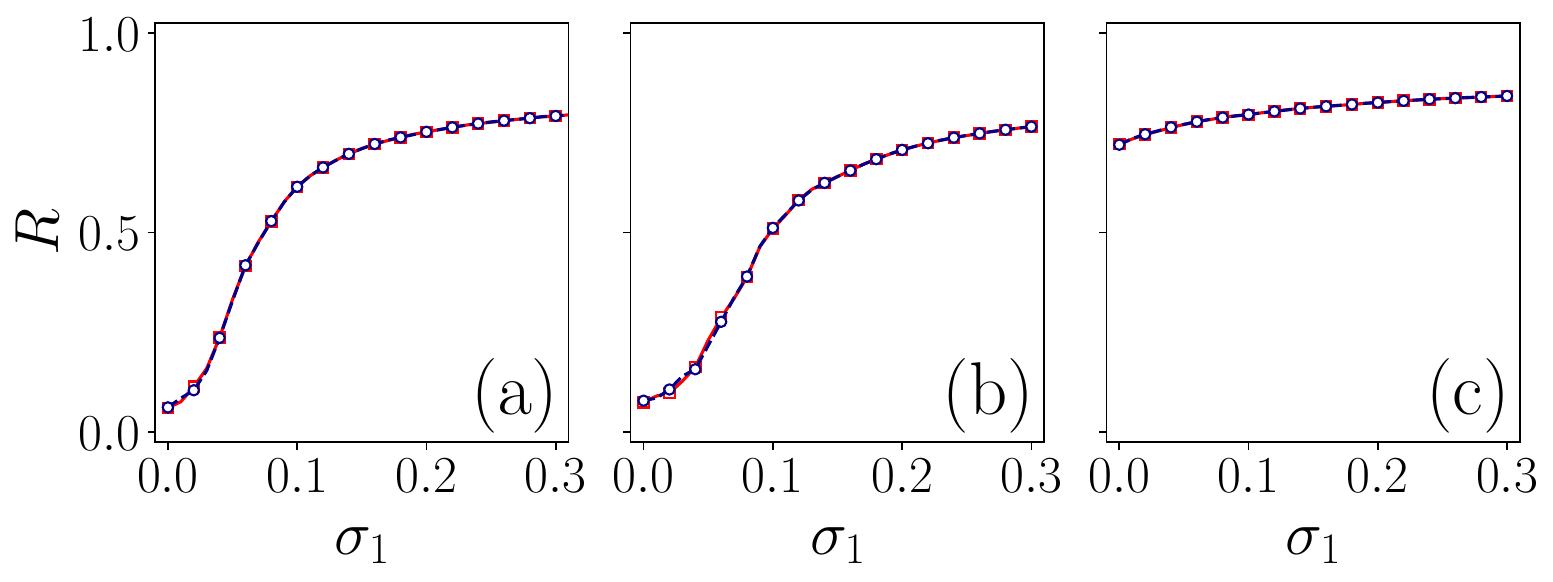}
\caption{The value of $R$ against $\sigma_1$ is calculated for the different values of (a) $\sigma_2 =0.0,$ (b) $\sigma_2 =0.1$ and (c) $\sigma_2 =0.3$ when all the links and triangles adapt competitively. As $\sigma_2$ increases, the system starts to partially synchronize even when there are no pairwise interactions. Since the presence of competitively adapting nodes in both links and triangles, for the same values of $\sigma_2 = 0.3$, as in the previous cases, the system does not exhibit complete synchrony.}
\label{fig:4}
\end{figure}

\begin{figure}[!ht]
\centering
\includegraphics[width=0.95\linewidth]{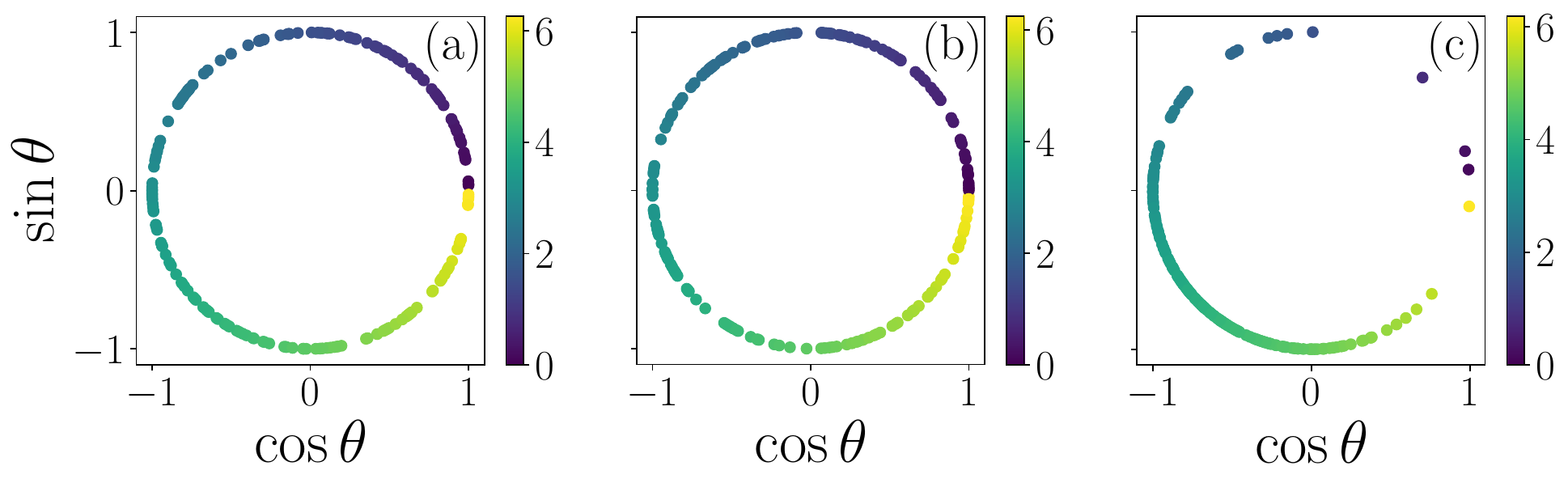}
\caption{The position of phases on unit circle for the different values of (a) $\sigma_2 = 0.0$, (b) $\sigma_2=0.1$, and (c) $\sigma_2=0.3$ when all the links and the triangles adapt competitively when $\sigma_1 = 0.0$. As $\sigma_2$ increases, the system starts to partially synchronize even when there are no pairwise interactions.}
\label{fig:5}
\end{figure}

Finally, we consider a case in which the nodes in both links and triangles adapt competitively. For that, we choose $ f_1 = f_2 = 1.0 $. As we see from Fig.~\ref{fig:4}(a), when there is no non-pairwise interaction (i.e., $\sigma_2=0.0)$ and the nodes in the links adapt competitively, the system goes into a continuous transition. If we include higher-order interaction term with $\sigma_2=0.05$, the initial $R$ increases as long as $\sigma_1 = 0.0$, and evidently, the system never goes to complete synchrony. It could be because the way to achieve the explosive transition is to avoid forming any clusters since the presence of competitive adaptation, the system forms an initial cluster and goes to synchrony as the size of the cluster increases. Figure~\ref{fig:4} shows $R$ versus $\sigma_1$ for various $\sigma_2$ computed numerically for different values of $\sigma_2 =0.0, 0.1, 0.3$. As we increase the values of $\sigma_2$, the system forms a cluster even when there are no pairwise interactions. To visualize this, we solve the dynamical equation when $\sigma_1 = 0.0$ for different values of $\sigma_2$ and we plot the phases over a unit circle and the results are shown in Fig.~\ref{fig:5}. The color bar denotes the phases of the oscillators, which are distributed as $[0, 2\pi]$. When $\sigma_2 = 0.0$, the phases of the oscillators distributed throughout the unit circle show an incoherent state [Fig.~\ref{fig:5}(a)]. As we increase $\sigma_2=0.1$, initially the state remains incoherent [Fig.~\ref{fig:5}(b)]. When the higher-order coupling strength is large enough, the system is partially synchronized with a cluster of oscillators having the same phase and a few drifting oscillators [Fig.~\ref{fig:5}(c)]. The formation of this cluster prevents explosive synchronization when the nodes adapt competitively. %

The above two cases demonstrate the possibility of synchronization even in the absence of pairwise interactions. So far, achieving synchronization in static networks has been thought to require pairwise interactions. However, our findings show that as in time-varying networks where the configuration of the system changes with time, can go to synchrony even when pairwise interactions are weak \cite{anwar2023synchronization}. Static networks can also exhibit synchronization when nodes adapt competitively and cooperatively with higher-order interactions. We can also see that cooperative adaptation is necessary for explosive synchronization. In the absence of cooperative adaptation, the system may go into synchrony, but the transition will not be explosive, even in presence of higher-order interactions. This scenario can be seen in epidemic spreading, where higher the number of people who are infected, the faster the spreading of the disease. This could be seen as a cooperatively adaptive dynamics, the presence of the higher-order interactions fastens the explosive synchronization, i.e., the system goes to a coherent state for the smaller values of $\sigma_1$, which means even in the case of less pairwise interactions or in the absence of pairwise interactions, the cooperative nodes in higher-order interactions cause the epidemic. Furthermore, as we can see from Fig.~\ref{fig:1} initially, the presence of triangles with competitively participating nodes promotes explosive synchronization. But, when the $\sigma_2$ is strong enough, the explosive synchronization vanishes. Thus, competitive nodes in links and triangles help sustain synchronization with a particular order parameter but prevent explosive transitions. Hence, the presence and absence of higher-order interactions determine when the system goes to synchronization, the presence of cooperative and competitive adaptation determines the way the system goes to synchronization, and our analytical results clearly support this.

\subsection{Effect on the number of nodes}
\begin{figure}[!ht]
\centering
\includegraphics[width=0.95\linewidth]{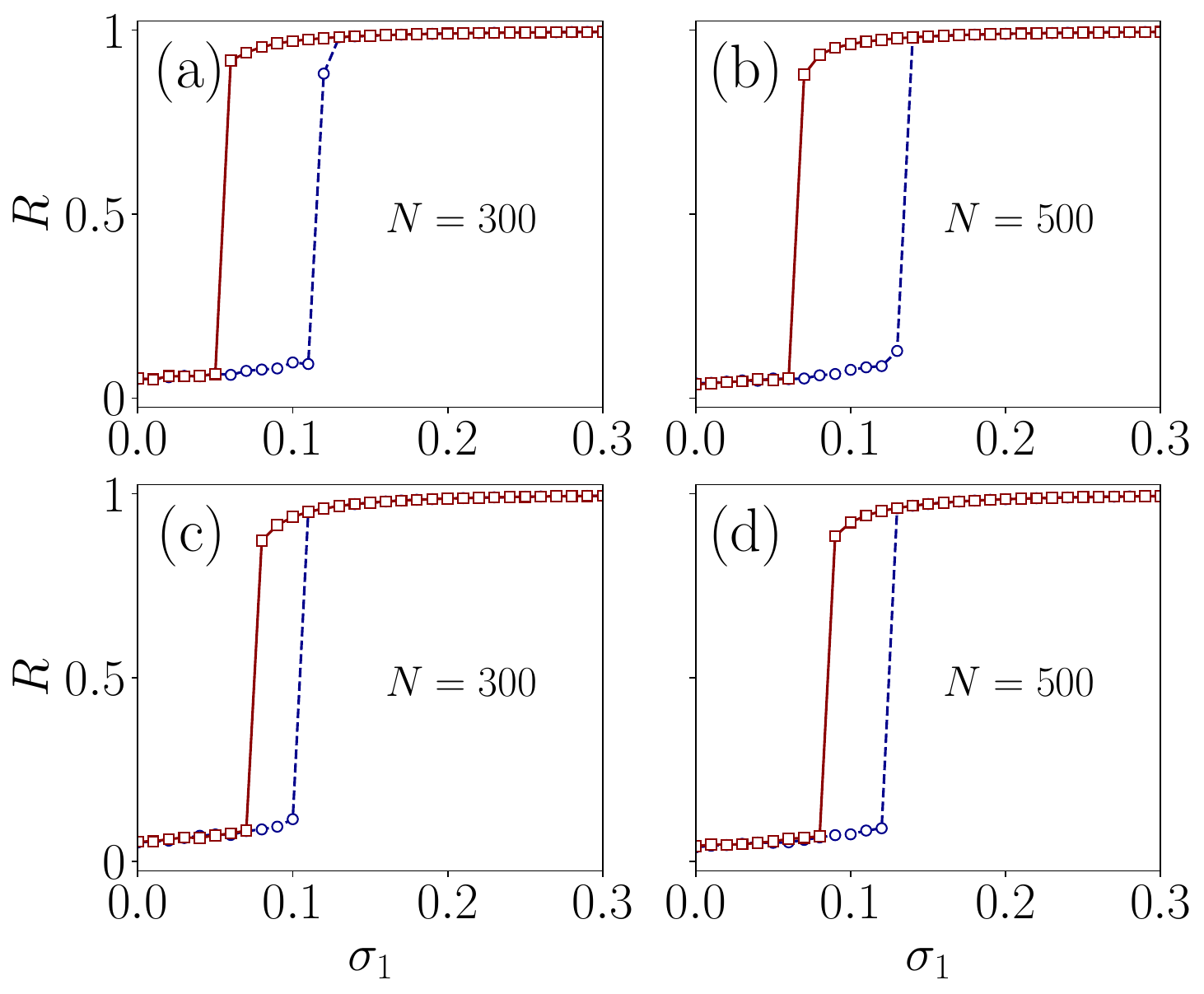}
\caption{Plots showing $R$ against $\sigma_1$ calculated for $K^{(1)} = 20$ and $\sigma_2 = 0.05$  for the different values of $N$: (a) $N= 300$  and  (b) $N = 500$ when nodes in the links and the triangles adapt cooperatively, and  (c) $N = 300$ and (d) $N= 500$  when nodes in the links and the triangles adapt cooperatively and competitively, respectively.}
\label{fig:6}
\end{figure}

Next, we analyze how the total number of oscillators in the system affects synchronization transitions. For this, we consider two cases: $(i)$ all nodes in links and triangles adapt cooperatively, and $(ii)$ pairwise links adapt cooperatively, and non-pairwise nodes adapt competitively. Keeping $\sigma_2 = 0.05$ and mean degree $K^{(1)}= 20$ fixed, we choose the number of oscillators $N=300$ and $N=500$. As the number of oscillators increases, stronger coupling is required to synchronize the system during forward transitions, resulting in more stable incoherent states. Conversely, on backward transitions, the system desynchronizes earlier if the system has a higher number of oscillators. We demonstrate these results in Fig.~\ref{fig:6}.

A higher number of oscillators with a fixed mean degree leads to sparser networks. In this context, a network with $N=300$ is denser than one with $N=500$. We can achieve faster synchronization with $N=500$ by increasing the mean degree. Therefore, the system's synchronization behavior depends not solely on the total number of oscillators but also on how densely or sparsely the nodes are connected.

Having analyzed the impact of network sparsity or density with cooperative adaptations only, we now focus on the scenario where all nodes within links adapt cooperatively, and all nodes within triangles adapt competitively. Using networks with $N = 300$ and $500$ oscillators and an average degree of $K^{(1)} = 20$, we observe that sparse networks exhibit a slight increase in hysteresis. Comparing these results to previous cases where both adaptations are cooperative, we find a decreased hysteresis width. For the other two adaptation scenarios (IV C and IV D), the qualitative behavior remains unchanged as the number of oscillators increases.

\subsection{Effect of the mean degree}
\begin{figure}[!ht]
    \centering
    \includegraphics[width=0.95\linewidth]{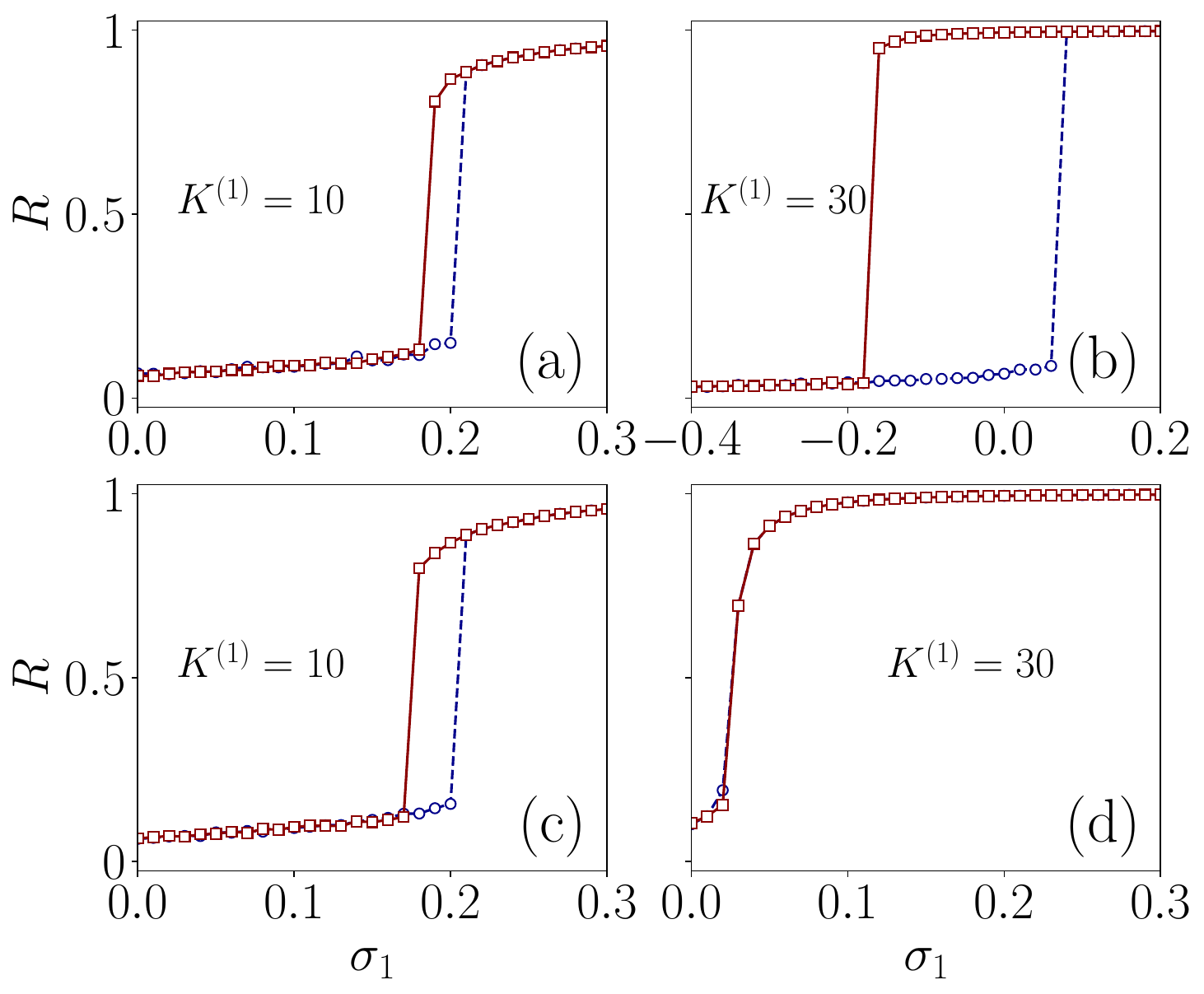}
    \caption{The plots of the global order parameter $R$ as a function of $\sigma_1$ calculated for $N = 200$ and $\sigma_2 = 0.05$  for   different mean degrees: (a) $K^{(1)}= 10$  and  (b) $K^{(1)}= 30$ when nodes in the links and the triangles adapt cooperatively. Similarly, in (c) $K^{(1)}= 10$ and  (d) $K^{(1)}= 30$  when nodes in the links and the triangles adapt cooperatively and competitively, respectively.  Denser networks increase explosive synchronization with widened hysteresis when higher-order interactions adapt cooperatively, and it hinders explosive synchronization when they adapt competitively.}
    \label{fig:7}
\end{figure}

In this subsection, we investigate the effects of mean degree on the  synchronization transitions. We consider $N=200$ and $\sigma_2 = 0.05$ and analyze the synchronization for two different values of mean degree, namely, $K^{(1)}= 10$ and $30$. When both adaptations are cooperative, the system exhibits an explosive synchronization transition. Increasing the mean degree accelerates the transition to synchronization. However, for sufficiently large enough mean degree, the system undergoes desynchronization during backward transitions for negative $\sigma_1$ values. Figures~\ref{fig:7}(a) and ~\ref{fig:7}(b) depict the variation of order parameter $R$ against $\sigma_1$ for (a) $K^{(1)}= 10$ and (b) $K^{(1)}= 30$, respectively. %

Finally, we consider the case when nodes in links adapt cooperatively while links in triangles adapt competitively. A network with a larger mean degree goes to synchronization non-explosively. In Figs.~\ref{fig:7}(c) and \ref{fig:7}(d), we show the variation of order parameter $R$ against $\sigma_1$ for (a) $K^{(1)}= 10$ and (b) $K^{(1)}= 30$, respectively. The system follows different paths to synchronization when nodes in higher-order interaction adapt cooperatively and competitively because of the larger number of higher-order interactions. As we can see, cooperative adaptation of nodes in higher-order interactions promotes explosive synchronization, and when the adaptation is competitive, the transition changes from first order to second order.

For the other two cases of adaptations, we can observe the same pattern as discussed for $N=200$ oscillators above. It is proven that cooperative adaptation is needed for the explosive transition and when competitive adaptation is strong, the system will not synchronize explosively. When the mean degree changes, the qualitative behavior is same. Higher mean degree and higher number of higher-order interactions strengthen the effects of higher-order adaptation.

\section{Analytical verification for  globally connected network}
\label{sec:4}
In order to verify our results, we analytically derive the expression for the global order parameter using the mean field approximation for the case of global coupling topology. The dynamical equation for the globally coupled Kuramoto oscillators with cooperative and competitive adaptations can be written by modifying Eq.~\eqref{eq:1} as, 
\begin{align}
\dot \phi_{i} = &\, \omega_i + \alpha {\sigma_1}\frac{1}{N}\sum_{j=1}^N  \sin(\phi_j - \phi_i ) \notag \\ 
& \, + \beta{\sigma_2} \frac{1}{N^2}\sum_{j=1}^N \sum_{p=1}^N  \sin(2 \phi_j - \phi_i - \phi_p ). \label{eq:2} 
\end{align}
Here, $\alpha$ and $\beta$ are the adaptation parameters. $\alpha = \beta = R$ for the cooperative adaptation and  $\alpha = \beta = 1-R$ for the competitive adaptation.
After using the mean-filed approximation, the above equation can be written as,
\begin{align}
\dot \phi_{i} = & \, \omega_i + \alpha{\sigma_1}R\sin(\psi - \phi_i ) 
+ \beta{\sigma_2}R r_c \sin(\psi_c - \psi - \phi_i ), \label{eq:3}
\end{align}
where $R$ and $r_c$ represent global and cluster order parameters, respectively. The $r_c$ is given by, 
\begin{align}
r_c \mathrm{e}^{\mathrm{i}\psi_c} & = \frac{1}{N} \sum_{j=1}^{N} \mathrm{e}^{\mathrm{i}2\phi_j}\label{eq:3a}
\end{align}
When $r_c=1$ and $R=0$, the oscillators form two distinct clusters and when $r_c=R=1$, the system is completely synchronized by forming one cluster  \cite{moyal2024rotating}.

\begin{figure}[!ht]
    \centering
    \includegraphics[width=0.95\linewidth]{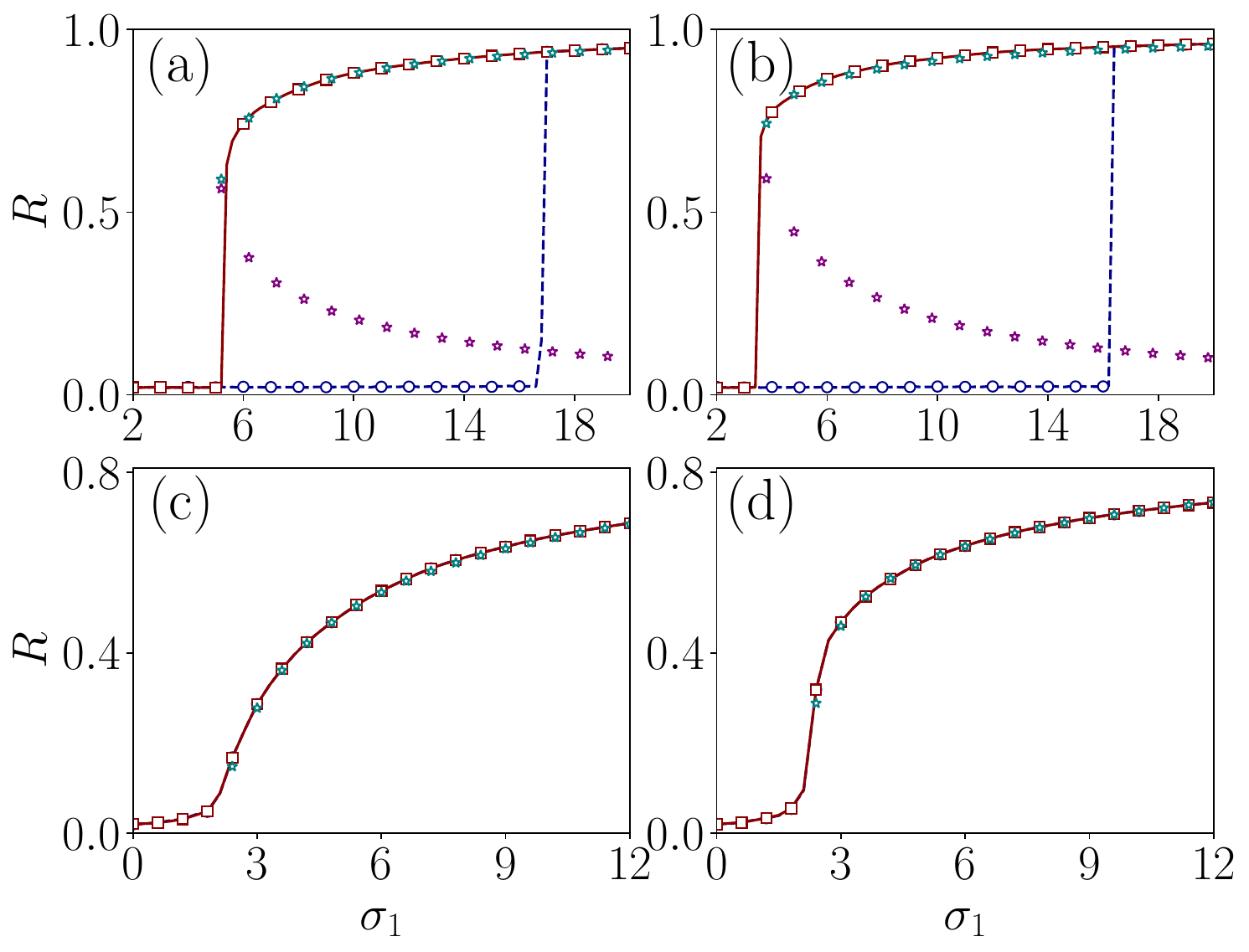}
    \caption{The variation of $R$ by changing $\sigma_1$ when both the pairwise and non-pairwise interactions adapt cooperatively for  (a) $\sigma_2 = 0.0$ and (b) $\sigma_2=4.0$ with $N=2000$. As $\sigma_2$ increases, the backward critical points shift to the lower values of the $\sigma_1$ and hence increase in the width of the hysteresis. Similarly, the variation of $R$ vs $\sigma_1$ when both the pairwise and non-pairwise interactions adapt competitively for (c) $\sigma_2 = 0.0$ and (d) $\sigma_2=8.0$. The markers star depict the analytical results and blue circles (red diamonds) are for forward (backward) transition  correspond to the numerical results. The presence and absence of two branches of solutions for $R \in [0,1]$  indicate the presence and absence of hysteresis.}
    \label{fig:8}
\end{figure}

In the continuum limit, i.e., $N \to \infty$, to reduce the dimension of the system of equations, we use Ott-Antonsen approach\cite{ott2008low} and we can write a periodic density function $f(\theta, \omega, t)$ that can describe the state of the system in that limit. Since it is a periodic function, it can be expanded into a Fourier series as,
\begin{align}
    f(\theta, \omega, t) = \frac{g(\omega)}{2\pi}[1+\sum_{n=1}^{\infty} f_n(\omega,t) \mathrm{e}^{\mathrm{i}n\theta} + \mathrm{c.c}]. \label{eq:4}
\end{align}
Here, $\mathrm{c.c}$ represents the complex conjugate and $f_n(\omega,t)$ is the $n$-th Fourier coefficient. The choice of the values of $f_n(\omega,t)$ can be taken from the Ott-Antonsen approach \cite{ott2008low} as,
\begin{align}
  f_n(\omega,t) = [\gamma(\omega,t)]^n,  \label{eq:4a}
\end{align}
with $\lvert \gamma(\omega,t) \rvert \leq 1$ and $g(\omega)$ is the Lorentzian frequency distribution represented as, 
\begin{align}
    g(\omega) = \frac{\Delta}{\pi[(\omega - \omega_0)^2 + \Delta^2]}. \label{eq:4b}
\end{align}
To get the expression for the global order parameter, we utilize the fact that since the total number of oscillators $N$ is conserved, the density function must satisfy the continuity equation,
 \begin{align}
     \frac{\partial f}{\partial t} + \frac{\partial (f\dot\phi)}{\partial \phi} = 0. \label{eq:5}
 \end{align}
After using the $f$ and $\dot\phi$ using equations \eqref{eq:3} and \eqref{eq:4}, the continuity equation can be written as in terms of $\gamma$ as
\begin{align}
    \dot\gamma = & -\mathrm{i} \omega\gamma + \frac{\sigma_1R\alpha}{2}\left(e^{-\mathrm{i}\psi_1} - \gamma^2e^{i\psi_1}\right) \notag \\ & + \frac{\sigma_2Rr_c\beta}{2}\left(\mathrm{e}^{-\mathrm{i}(\psi_2 - \psi_1)} - \gamma^2\mathrm{e}^{\mathrm{i}(\psi_2 - \psi_1)}\right).\label{eq:6}
\end{align} 
To further simplify this equation, we adopt the relation between $R$, $r_c$ and $\gamma$ as \cite{sharma2024synchronization}, $R \mathrm{e}^{-\mathrm{i} \psi_1} = \gamma(\omega_0 - \mathrm{i}\Delta, t)$ and $r_ce^{-\mathrm{i}\psi_2} =  \left(\gamma\left(\omega_0 - \mathrm{i}\Delta, t \right)\right)^2$.  Equating the real part of  Eq.~(\ref{eq:6}) and considering $\dot R = 0$ gives us the fixed point as 
\begin{align}
 R = \frac{\sigma_1R\alpha}{2} \left(1 - R^2\right) + \frac{\sigma_2R^3\beta}{2}\left(1 - R^2\right). \label{eq:7}
\end{align}
To verify the results, we first consider the case in which the nodes in both links and triangles adapt cooperatively by setting \(\alpha = \beta = 1\).  The numerical and analytical results are presented in Fig. \ref{fig:8}. We consider two values for \(\sigma_2\): \(0\) and \(4\). As we increase \(\sigma_2\), the forward critical point shifts to higher values of \(\sigma_1\), while the backward critical point moves to lower values of \(\sigma_1\). We use Eq.~\eqref{eq:3} for the numerical analysis for \(N = 2000\). 

Next, we analyze the case where both interactions adapt competitively. For this, we choose the values of \(\sigma_2\) as \(0\) and \(8\). As from the figure, we can observe that when both the pairwise and non-pairwise interactions adapt cooperatively, we have two branches of non-trivial solutions fall in the range of $[0, 1]$  when we solve Eq.~\eqref{eq:7} which proves the presence of the hysteresis correspond to the explosive synchronization [see Figs. \ref{fig:8}(a) and \ref{fig:8}(b)]. Conversely, when both interactions adapt competitively, we have a single branch of solution which demonstrate the second order transition [see Figs. \ref{fig:8}(c) and \ref{fig:8}(d)]. From the results, it is evident that the cooperative adaptations promote the first order synchronization while competitive adaptations may force the system to undergo second order transitions. In both cases, the higher-order interactions aid faster synchronization with wider hysteresis. Using Eq.~\eqref{eq:7}, we can conduct preliminary tests to identify the type of synchronization transition in large complex systems for a specific adaptation without solving the dynamical equations, as demonstrated in Ref.~\onlinecite{Jenifer2024}.

\section{Conclusion}
\label{sec:5}
We investigated the interplay between cooperation, competition, and higher-order interactions on synchronization in complex networks. Our findings indicate that pure competition among elements may inhibit explosive synchronization, while cooperation can facilitate it. Interestingly, higher-order interactions, such as three-body interactions, can even result in explosive synchronization under specific conditions. In other scenarios, higher-order interactions can cause the system to undergo a phase transition, from first-order to second-order, and continuous to discontinuous transitions. The presence of higher-order interactions, combined with cooperative and competitive adaptation, not only allows us to control when a system synchronizes but also how it synchronizes, which includes achieving both continuous, sharp, and discontinuous transitions. Due to the cooperative and competitive dynamics in the higher-order networks, the system can achieve synchronization even in the complete absence of pairwise interactions in static networks. This finding opens the question of whether a carefully balanced mix of competition and cooperation might be beneficial for achieving optimal synchronization. We also investigated the effects of the total number of oscillators and mean degrees. It is observed that the qualitative behaviors remains the same with changes in the transition points. In denser networks, the effect of the adaptations is prominent. Moreover, we supported our findings analytically using globally coupled network topology. Future studies could explore how the interaction between these forces affects synchronization in different systems, potentially revealing scenarios where a controlled level of competition can improve synchronization efficiency.

\acknowledgments
The work of S.N.J. and P.M. is supported by MoE RUSA 2.0 (Bharathidasan University-Physical Sciences).

\section*{Data availability}
The data supporting this study's findings are available from the corresponding author upon reasonable request. 

\section*{References}

\begin{thebibliography}{27}%
\makeatletter
\providecommand \@ifxundefined [1]{%
 \@ifx{#1\undefined}
}%
\providecommand \@ifnum [1]{%
 \ifnum #1\expandafter \@firstoftwo
 \else \expandafter \@secondoftwo
 \fi
}%
\providecommand \@ifx [1]{%
 \ifx #1\expandafter \@firstoftwo
 \else \expandafter \@secondoftwo
 \fi
}%
\providecommand \natexlab [1]{#1}%
\providecommand \enquote  [1]{``#1''}%
\providecommand \bibnamefont  [1]{#1}%
\providecommand \bibfnamefont [1]{#1}%
\providecommand \citenamefont [1]{#1}%
\providecommand \href@noop [0]{\@secondoftwo}%
\providecommand \href [0]{\begingroup \@sanitize@url \@href}%
\providecommand \@href[1]{\@@startlink{#1}\@@href}%
\providecommand \@@href[1]{\endgroup#1\@@endlink}%
\providecommand \@sanitize@url [0]{\catcode `\\12\catcode `\$12\catcode `\&12\catcode `\#12\catcode `\^12\catcode `\_12\catcode `\%12\relax}%
\providecommand \@@startlink[1]{}%
\providecommand \@@endlink[0]{}%
\providecommand \url  [0]{\begingroup\@sanitize@url \@url }%
\providecommand \@url [1]{\endgroup\@href {#1}{\urlprefix }}%
\providecommand \urlprefix  [0]{URL }%
\providecommand \Eprint [0]{\href }%
\providecommand \doibase [0]{https://doi.org/}%
\providecommand \selectlanguage [0]{\@gobble}%
\providecommand \bibinfo  [0]{\@secondoftwo}%
\providecommand \bibfield  [0]{\@secondoftwo}%
\providecommand \translation [1]{[#1]}%
\providecommand \BibitemOpen [0]{}%
\providecommand \bibitemStop [0]{}%
\providecommand \bibitemNoStop [0]{.\EOS\space}%
\providecommand \EOS [0]{\spacefactor3000\relax}%
\providecommand \BibitemShut  [1]{\csname bibitem#1\endcsname}%
\let\auto@bib@innerbib\@empty
\bibitem [{\citenamefont {Battiston}\ \emph {et~al.}(2020)\citenamefont {Battiston}, \citenamefont {Cencetti}, \citenamefont {Iacopini}, \citenamefont {Latora}, \citenamefont {Lucas}, \citenamefont {Patania}, \citenamefont {Young},\ and\ \citenamefont {Petri}}]{Battiston2020}%
  \BibitemOpen
  \bibfield  {author} {\bibinfo {author} {\bibfnamefont {F.}~\bibnamefont {Battiston}}, \bibinfo {author} {\bibfnamefont {G.}~\bibnamefont {Cencetti}}, \bibinfo {author} {\bibfnamefont {I.}~\bibnamefont {Iacopini}}, \bibinfo {author} {\bibfnamefont {V.}~\bibnamefont {Latora}}, \bibinfo {author} {\bibfnamefont {M.}~\bibnamefont {Lucas}}, \bibinfo {author} {\bibfnamefont {A.}~\bibnamefont {Patania}}, \bibinfo {author} {\bibfnamefont {J.-G.}\ \bibnamefont {Young}},\ and\ \bibinfo {author} {\bibfnamefont {G.}~\bibnamefont {Petri}},\ }\bibfield  {title} {\enquote {\bibinfo {title} {Networks beyond pairwise interactions: structure and dynamics},}\ }\href {https://doi.org/10.1016/j.physrep.2020.05.004} {\bibfield  {journal} {\bibinfo  {journal} {Phys. Rep.}\ }\textbf {\bibinfo {volume} {874}},\ \bibinfo {pages} {1--92} (\bibinfo {year} {2020})}\BibitemShut {NoStop}%
\bibitem [{\citenamefont {Battiston}\ \emph {et~al.}(2021)\citenamefont {Battiston}, \citenamefont {Amico}, \citenamefont {Barrat}, \citenamefont {Bianconi}, \citenamefont {Ferraz~de Arruda}, \citenamefont {Franceschiello}, \citenamefont {Iacopini}, \citenamefont {K{\'e}fi}, \citenamefont {Latora}, \citenamefont {Moreno} \emph {et~al.}}]{Battiston2021}%
  \BibitemOpen
  \bibfield  {author} {\bibinfo {author} {\bibfnamefont {F.}~\bibnamefont {Battiston}}, \bibinfo {author} {\bibfnamefont {E.}~\bibnamefont {Amico}}, \bibinfo {author} {\bibfnamefont {A.}~\bibnamefont {Barrat}}, \bibinfo {author} {\bibfnamefont {G.}~\bibnamefont {Bianconi}}, \bibinfo {author} {\bibfnamefont {G.}~\bibnamefont {Ferraz~de Arruda}}, \bibinfo {author} {\bibfnamefont {B.}~\bibnamefont {Franceschiello}}, \bibinfo {author} {\bibfnamefont {I.}~\bibnamefont {Iacopini}}, \bibinfo {author} {\bibfnamefont {S.}~\bibnamefont {K{\'e}fi}}, \bibinfo {author} {\bibfnamefont {V.}~\bibnamefont {Latora}}, \bibinfo {author} {\bibfnamefont {Y.}~\bibnamefont {Moreno}}, \emph {et~al.},\ }\bibfield  {title} {\enquote {\bibinfo {title} {The physics of higher-order interactions in complex systems},}\ }\href {https://doi.org/10.1038/s41567-021-01371-4} {\bibfield  {journal} {\bibinfo  {journal} {Nat. Phys.}\ }\textbf {\bibinfo {volume} {17}},\ \bibinfo {pages} {1093--1098} (\bibinfo {year} {2021})}\BibitemShut {NoStop}%
\bibitem [{\citenamefont {Majhi}, \citenamefont {Perc},\ and\ \citenamefont {Ghosh}(2022)}]{Majhi2022}%
  \BibitemOpen
  \bibfield  {author} {\bibinfo {author} {\bibfnamefont {S.}~\bibnamefont {Majhi}}, \bibinfo {author} {\bibfnamefont {M.}~\bibnamefont {Perc}},\ and\ \bibinfo {author} {\bibfnamefont {D.}~\bibnamefont {Ghosh}},\ }\bibfield  {title} {\enquote {\bibinfo {title} {Dynamics on higher-order networks: A review},}\ }\href {https://doi.org/10.1098/rsif.2022.0043} {\bibfield  {journal} {\bibinfo  {journal} {J. R. Soc. Interface}\ }\textbf {\bibinfo {volume} {19}},\ \bibinfo {pages} {20220043} (\bibinfo {year} {2022})}\BibitemShut {NoStop}%
\bibitem [{\citenamefont {Iacopini}\ \emph {et~al.}(2019)\citenamefont {Iacopini}, \citenamefont {Petri}, \citenamefont {Barrat},\ and\ \citenamefont {Latora}}]{Iacopini2019}%
  \BibitemOpen
  \bibfield  {author} {\bibinfo {author} {\bibfnamefont {I.}~\bibnamefont {Iacopini}}, \bibinfo {author} {\bibfnamefont {G.}~\bibnamefont {Petri}}, \bibinfo {author} {\bibfnamefont {A.}~\bibnamefont {Barrat}},\ and\ \bibinfo {author} {\bibfnamefont {V.}~\bibnamefont {Latora}},\ }\bibfield  {title} {\enquote {\bibinfo {title} {Simplicial models of social contagion},}\ }\href {https://doi.org/10.1038/s41467-019-10431-6} {\bibfield  {journal} {\bibinfo  {journal} {Nat. Commun.}\ }\textbf {\bibinfo {volume} {10}},\ \bibinfo {pages} {1--9} (\bibinfo {year} {2019})}\BibitemShut {NoStop}%
\bibitem [{\citenamefont {Boccaletti}\ \emph {et~al.}(2006)\citenamefont {Boccaletti}, \citenamefont {Latora}, \citenamefont {Moreno}, \citenamefont {Chavez},\ and\ \citenamefont {Hwang}}]{Boccaletti2006}%
  \BibitemOpen
  \bibfield  {author} {\bibinfo {author} {\bibfnamefont {S.}~\bibnamefont {Boccaletti}}, \bibinfo {author} {\bibfnamefont {V.}~\bibnamefont {Latora}}, \bibinfo {author} {\bibfnamefont {Y.}~\bibnamefont {Moreno}}, \bibinfo {author} {\bibfnamefont {M.}~\bibnamefont {Chavez}},\ and\ \bibinfo {author} {\bibfnamefont {D.-U.}\ \bibnamefont {Hwang}},\ }\bibfield  {title} {\enquote {\bibinfo {title} {Complex networks: Structure and dynamics},}\ }\href {https://doi.org/10.1016/j.physrep.2005.10.009} {\bibfield  {journal} {\bibinfo  {journal} {Phys. Rep.}\ }\textbf {\bibinfo {volume} {424}},\ \bibinfo {pages} {175--308} (\bibinfo {year} {2006})}\BibitemShut {NoStop}%
\bibitem [{\citenamefont {Sawicki}\ \emph {et~al.}(2023)\citenamefont {Sawicki}, \citenamefont {Berner}, \citenamefont {Loos}, \citenamefont {Anvari}, \citenamefont {Bader}, \citenamefont {Barfuss}, \citenamefont {Botta}, \citenamefont {Brede}, \citenamefont {Franovi{\'c}}, \citenamefont {Gauthier} \emph {et~al.}}]{sawicki2023perspectives}%
  \BibitemOpen
  \bibfield  {author} {\bibinfo {author} {\bibfnamefont {J.}~\bibnamefont {Sawicki}}, \bibinfo {author} {\bibfnamefont {R.}~\bibnamefont {Berner}}, \bibinfo {author} {\bibfnamefont {S.~A.}\ \bibnamefont {Loos}}, \bibinfo {author} {\bibfnamefont {M.}~\bibnamefont {Anvari}}, \bibinfo {author} {\bibfnamefont {R.}~\bibnamefont {Bader}}, \bibinfo {author} {\bibfnamefont {W.}~\bibnamefont {Barfuss}}, \bibinfo {author} {\bibfnamefont {N.}~\bibnamefont {Botta}}, \bibinfo {author} {\bibfnamefont {N.}~\bibnamefont {Brede}}, \bibinfo {author} {\bibfnamefont {I.}~\bibnamefont {Franovi{\'c}}}, \bibinfo {author} {\bibfnamefont {D.~J.}\ \bibnamefont {Gauthier}}, \emph {et~al.},\ }\bibfield  {title} {\enquote {\bibinfo {title} {Perspectives on adaptive dynamical systems},}\ }\href {https://doi.org/10.1063/5.0147231} {\bibfield  {journal} {\bibinfo  {journal} {Chaos}\ }\textbf {\bibinfo {volume} {33}},\ \bibinfo {pages} {071501} (\bibinfo {year} {2023})}\BibitemShut {NoStop}%
\bibitem [{\citenamefont {Berner}\ \emph {et~al.}(2023)\citenamefont {Berner}, \citenamefont {Gross}, \citenamefont {Kuehn}, \citenamefont {Kurths},\ and\ \citenamefont {Yanchuk}}]{berner2023adaptive}%
  \BibitemOpen
  \bibfield  {author} {\bibinfo {author} {\bibfnamefont {R.}~\bibnamefont {Berner}}, \bibinfo {author} {\bibfnamefont {T.}~\bibnamefont {Gross}}, \bibinfo {author} {\bibfnamefont {C.}~\bibnamefont {Kuehn}}, \bibinfo {author} {\bibfnamefont {J.}~\bibnamefont {Kurths}},\ and\ \bibinfo {author} {\bibfnamefont {S.}~\bibnamefont {Yanchuk}},\ }\bibfield  {title} {\enquote {\bibinfo {title} {Adaptive dynamical networks},}\ }\href {https://doi.org/10.1016/j.physrep.2023.08.001} {\bibfield  {journal} {\bibinfo  {journal} {Phys. Rep.}\ }\textbf {\bibinfo {volume} {1031}},\ \bibinfo {pages} {1--59} (\bibinfo {year} {2023})}\BibitemShut {NoStop}%
\bibitem [{\citenamefont {Kasatkin}\ \emph {et~al.}(2017)\citenamefont {Kasatkin}, \citenamefont {Yanchuk}, \citenamefont {Sch\"oll},\ and\ \citenamefont {Nekorkin}}]{kasatkin2017self}%
  \BibitemOpen
  \bibfield  {author} {\bibinfo {author} {\bibfnamefont {D.~V.}\ \bibnamefont {Kasatkin}}, \bibinfo {author} {\bibfnamefont {S.}~\bibnamefont {Yanchuk}}, \bibinfo {author} {\bibfnamefont {E.}~\bibnamefont {Sch\"oll}},\ and\ \bibinfo {author} {\bibfnamefont {V.~I.}\ \bibnamefont {Nekorkin}},\ }\bibfield  {title} {\enquote {\bibinfo {title} {Self-organized emergence of multilayer structure and chimera states in dynamical networks with adaptive couplings},}\ }\href {https://doi.org/10.1103/PhysRevE.96.062211} {\bibfield  {journal} {\bibinfo  {journal} {Phys. Rev. E}\ }\textbf {\bibinfo {volume} {96}},\ \bibinfo {pages} {062211} (\bibinfo {year} {2017})}\BibitemShut {NoStop}%
\bibitem [{\citenamefont {Berner}, \citenamefont {Scholl},\ and\ \citenamefont {Yanchuk}(2019)}]{berner2019multiclusters}%
  \BibitemOpen
  \bibfield  {author} {\bibinfo {author} {\bibfnamefont {R.}~\bibnamefont {Berner}}, \bibinfo {author} {\bibfnamefont {E.}~\bibnamefont {Scholl}},\ and\ \bibinfo {author} {\bibfnamefont {S.}~\bibnamefont {Yanchuk}},\ }\bibfield  {title} {\enquote {\bibinfo {title} {Multiclusters in networks of adaptively coupled phase oscillators},}\ }\href {https://doi.org/10.1137/18M1210150} {\bibfield  {journal} {\bibinfo  {journal} {SIAM J. Appl. Dyn. Syst.}\ }\textbf {\bibinfo {volume} {18}},\ \bibinfo {pages} {2227--2266} (\bibinfo {year} {2019})}\BibitemShut {NoStop}%
\bibitem [{\citenamefont {Berner}\ \emph {et~al.}(2021)\citenamefont {Berner}, \citenamefont {Vock}, \citenamefont {Sch\"oll},\ and\ \citenamefont {Yanchuk}}]{berner2021desynchronization}%
  \BibitemOpen
  \bibfield  {author} {\bibinfo {author} {\bibfnamefont {R.}~\bibnamefont {Berner}}, \bibinfo {author} {\bibfnamefont {S.}~\bibnamefont {Vock}}, \bibinfo {author} {\bibfnamefont {E.}~\bibnamefont {Sch\"oll}},\ and\ \bibinfo {author} {\bibfnamefont {S.}~\bibnamefont {Yanchuk}},\ }\bibfield  {title} {\enquote {\bibinfo {title} {Desynchronization transitions in adaptive networks},}\ }\href {https://doi.org/10.1103/PhysRevLett.126.028301} {\bibfield  {journal} {\bibinfo  {journal} {Phys. Rev. Lett.}\ }\textbf {\bibinfo {volume} {126}},\ \bibinfo {pages} {028301} (\bibinfo {year} {2021})}\BibitemShut {NoStop}%
\bibitem [{\citenamefont {Liang}(2021)}]{liang2021measuring}%
  \BibitemOpen
  \bibfield  {author} {\bibinfo {author} {\bibfnamefont {X.~S.}\ \bibnamefont {Liang}},\ }\bibfield  {title} {\enquote {\bibinfo {title} {Measuring the importance of individual units in producing the collective behavior of a complex network},}\ }\href {https://doi.org/10.1063/5.0055051} {\bibfield  {journal} {\bibinfo  {journal} {Chaos}\ }\textbf {\bibinfo {volume} {31}},\ \bibinfo {pages} {093123} (\bibinfo {year} {2021})}\BibitemShut {NoStop}%
\bibitem [{\citenamefont {Winfree}(1967)}]{winfree1967biological}%
  \BibitemOpen
  \bibfield  {author} {\bibinfo {author} {\bibfnamefont {A.~T.}\ \bibnamefont {Winfree}},\ }\bibfield  {title} {\enquote {\bibinfo {title} {Biological rhythms and the behavior of populations of coupled oscillators},}\ }\href {https://doi.org/10.1016/0022-5193(67)90051-3} {\bibfield  {journal} {\bibinfo  {journal} {J. Theor. Biol.}\ }\textbf {\bibinfo {volume} {16}},\ \bibinfo {pages} {15--42} (\bibinfo {year} {1967})}\BibitemShut {NoStop}%
\bibitem [{\citenamefont {Kuramoto}(1975)}]{Kuramoto1975}%
  \BibitemOpen
  \bibfield  {author} {\bibinfo {author} {\bibfnamefont {Y.}~\bibnamefont {Kuramoto}},\ }\bibfield  {title} {\enquote {\bibinfo {title} {Self-entrainment of a population of coupled non-linear oscillators},}\ }in\ \href {https://doi.org/10.1007/BFb0013365} {\emph {\bibinfo {booktitle} {International symposium on mathematical problems in theoretical physics}}}\ (\bibinfo {organization} {Springer},\ \bibinfo {year} {1975})\ pp.\ \bibinfo {pages} {420--422}\BibitemShut {NoStop}%
\bibitem [{\citenamefont {Rodrigues}\ \emph {et~al.}(2016)\citenamefont {Rodrigues}, \citenamefont {Peron}, \citenamefont {Ji},\ and\ \citenamefont {Kurths}}]{rodrigues2016kuramoto}%
  \BibitemOpen
  \bibfield  {author} {\bibinfo {author} {\bibfnamefont {F.~A.}\ \bibnamefont {Rodrigues}}, \bibinfo {author} {\bibfnamefont {T.~K.~D.}\ \bibnamefont {Peron}}, \bibinfo {author} {\bibfnamefont {P.}~\bibnamefont {Ji}},\ and\ \bibinfo {author} {\bibfnamefont {J.}~\bibnamefont {Kurths}},\ }\bibfield  {title} {\enquote {\bibinfo {title} {The {K}uramoto model in complex networks},}\ }\href {https://doi.org/10.1016/j.physrep.2015.10.008} {\bibfield  {journal} {\bibinfo  {journal} {Phys. Rep.}\ }\textbf {\bibinfo {volume} {610}},\ \bibinfo {pages} {1--98} (\bibinfo {year} {2016})}\BibitemShut {NoStop}%
\bibitem [{\citenamefont {Arenas}\ \emph {et~al.}(2008)\citenamefont {Arenas}, \citenamefont {D{\'\i}az-Guilera}, \citenamefont {Kurths}, \citenamefont {Moreno},\ and\ \citenamefont {Zhou}}]{arenas2008synchronization}%
  \BibitemOpen
  \bibfield  {author} {\bibinfo {author} {\bibfnamefont {A.}~\bibnamefont {Arenas}}, \bibinfo {author} {\bibfnamefont {A.}~\bibnamefont {D{\'\i}az-Guilera}}, \bibinfo {author} {\bibfnamefont {J.}~\bibnamefont {Kurths}}, \bibinfo {author} {\bibfnamefont {Y.}~\bibnamefont {Moreno}},\ and\ \bibinfo {author} {\bibfnamefont {C.}~\bibnamefont {Zhou}},\ }\bibfield  {title} {\enquote {\bibinfo {title} {Synchronization in complex networks},}\ }\href {https://doi.org/10.1016/j.physrep.2008.09.002} {\bibfield  {journal} {\bibinfo  {journal} {Phys. Rep.}\ }\textbf {\bibinfo {volume} {469}},\ \bibinfo {pages} {93--153} (\bibinfo {year} {2008})}\BibitemShut {NoStop}%
\bibitem [{\citenamefont {G\'omez-Garde\~nes}, \citenamefont {Moreno},\ and\ \citenamefont {Arenas}(2007)}]{gomez2007paths}%
  \BibitemOpen
  \bibfield  {author} {\bibinfo {author} {\bibfnamefont {J.}~\bibnamefont {G\'omez-Garde\~nes}}, \bibinfo {author} {\bibfnamefont {Y.}~\bibnamefont {Moreno}},\ and\ \bibinfo {author} {\bibfnamefont {A.}~\bibnamefont {Arenas}},\ }\bibfield  {title} {\enquote {\bibinfo {title} {Paths to synchronization on complex networks},}\ }\href {https://doi.org/10.1103/PhysRevLett.98.034101} {\bibfield  {journal} {\bibinfo  {journal} {Phys. Rev. Lett.}\ }\textbf {\bibinfo {volume} {98}},\ \bibinfo {pages} {034101} (\bibinfo {year} {2007})}\BibitemShut {NoStop}%
\bibitem [{\citenamefont {G\'omez-Garde\~nes}\ \emph {et~al.}(2011)\citenamefont {G\'omez-Garde\~nes}, \citenamefont {G\'omez}, \citenamefont {Arenas},\ and\ \citenamefont {Moreno}}]{gomez2011explosive}%
  \BibitemOpen
  \bibfield  {author} {\bibinfo {author} {\bibfnamefont {J.}~\bibnamefont {G\'omez-Garde\~nes}}, \bibinfo {author} {\bibfnamefont {S.}~\bibnamefont {G\'omez}}, \bibinfo {author} {\bibfnamefont {A.}~\bibnamefont {Arenas}},\ and\ \bibinfo {author} {\bibfnamefont {Y.}~\bibnamefont {Moreno}},\ }\bibfield  {title} {\enquote {\bibinfo {title} {Explosive synchronization transitions in scale-free networks},}\ }\href {https://doi.org/10.1103/PhysRevLett.106.128701} {\bibfield  {journal} {\bibinfo  {journal} {Phys. Rev. Lett.}\ }\textbf {\bibinfo {volume} {106}},\ \bibinfo {pages} {128701} (\bibinfo {year} {2011})}\BibitemShut {NoStop}%
\bibitem [{\citenamefont {Zhang}\ \emph {et~al.}(2013)\citenamefont {Zhang}, \citenamefont {Hu}, \citenamefont {Kurths},\ and\ \citenamefont {Liu}}]{zhang2013explosive}%
  \BibitemOpen
  \bibfield  {author} {\bibinfo {author} {\bibfnamefont {X.}~\bibnamefont {Zhang}}, \bibinfo {author} {\bibfnamefont {X.}~\bibnamefont {Hu}}, \bibinfo {author} {\bibfnamefont {J.}~\bibnamefont {Kurths}},\ and\ \bibinfo {author} {\bibfnamefont {Z.}~\bibnamefont {Liu}},\ }\bibfield  {title} {\enquote {\bibinfo {title} {Explosive synchronization in a general complex network},}\ }\href {https://doi.org/10.1103/PhysRevE.88.010802} {\bibfield  {journal} {\bibinfo  {journal} {Phys. Rev. E}\ }\textbf {\bibinfo {volume} {88}},\ \bibinfo {pages} {010802} (\bibinfo {year} {2013})}\BibitemShut {NoStop}%
\bibitem [{\citenamefont {Zhang}\ \emph {et~al.}(2015)\citenamefont {Zhang}, \citenamefont {Boccaletti}, \citenamefont {Guan},\ and\ \citenamefont {Liu}}]{zhang2015explosive}%
  \BibitemOpen
  \bibfield  {author} {\bibinfo {author} {\bibfnamefont {X.}~\bibnamefont {Zhang}}, \bibinfo {author} {\bibfnamefont {S.}~\bibnamefont {Boccaletti}}, \bibinfo {author} {\bibfnamefont {S.}~\bibnamefont {Guan}},\ and\ \bibinfo {author} {\bibfnamefont {Z.}~\bibnamefont {Liu}},\ }\bibfield  {title} {\enquote {\bibinfo {title} {Explosive synchronization in adaptive and multilayer networks},}\ }\href {https://doi.org/10.1103/PhysRevLett.114.038701} {\bibfield  {journal} {\bibinfo  {journal} {Phys. Rev. Lett.}\ }\textbf {\bibinfo {volume} {114}},\ \bibinfo {pages} {038701} (\bibinfo {year} {2015})}\BibitemShut {NoStop}%
\bibitem [{\citenamefont {Frolov}\ \emph {et~al.}(2021)\citenamefont {Frolov}, \citenamefont {Rakshit}, \citenamefont {Maksimenko}, \citenamefont {Kirsanov}, \citenamefont {Ghosh},\ and\ \citenamefont {Hramov}}]{frolov2021coexistence}%
  \BibitemOpen
  \bibfield  {author} {\bibinfo {author} {\bibfnamefont {N.}~\bibnamefont {Frolov}}, \bibinfo {author} {\bibfnamefont {S.}~\bibnamefont {Rakshit}}, \bibinfo {author} {\bibfnamefont {V.}~\bibnamefont {Maksimenko}}, \bibinfo {author} {\bibfnamefont {D.}~\bibnamefont {Kirsanov}}, \bibinfo {author} {\bibfnamefont {D.}~\bibnamefont {Ghosh}},\ and\ \bibinfo {author} {\bibfnamefont {A.}~\bibnamefont {Hramov}},\ }\bibfield  {title} {\enquote {\bibinfo {title} {Coexistence of interdependence and competition in adaptive multilayer network},}\ }\href {https://doi.org/10.1016/j.chaos.2021.110955} {\bibfield  {journal} {\bibinfo  {journal} {Chaos, Solitons Fract.}\ }\textbf {\bibinfo {volume} {147}},\ \bibinfo {pages} {110955} (\bibinfo {year} {2021})}\BibitemShut {NoStop}%
\bibitem [{\citenamefont {Frolov}\ and\ \citenamefont {Hramov}(2021)}]{Frolov2021}%
  \BibitemOpen
  \bibfield  {author} {\bibinfo {author} {\bibfnamefont {N.}~\bibnamefont {Frolov}}\ and\ \bibinfo {author} {\bibfnamefont {A.}~\bibnamefont {Hramov}},\ }\bibfield  {title} {\enquote {\bibinfo {title} {{Extreme synchronization events in a Kuramoto model: The interplay between resource constraints and explosive transitions}},}\ }\href {https://doi.org/10.1063/5.0055156} {\bibfield  {journal} {\bibinfo  {journal} {Chaos}\ }\textbf {\bibinfo {volume} {31}},\ \bibinfo {pages} {063103} (\bibinfo {year} {2021})}\BibitemShut {NoStop}%
\bibitem [{\citenamefont {Frolov}\ and\ \citenamefont {Hramov}(2022)}]{frolov2022self}%
  \BibitemOpen
  \bibfield  {author} {\bibinfo {author} {\bibfnamefont {N.}~\bibnamefont {Frolov}}\ and\ \bibinfo {author} {\bibfnamefont {A.}~\bibnamefont {Hramov}},\ }\bibfield  {title} {\enquote {\bibinfo {title} {Self-organized bistability on scale-free networks},}\ }\href {https://doi.org/10.1103/PhysRevE.106.044301} {\bibfield  {journal} {\bibinfo  {journal} {Phys. Rev. E}\ }\textbf {\bibinfo {volume} {106}},\ \bibinfo {pages} {044301} (\bibinfo {year} {2022})}\BibitemShut {NoStop}%
\bibitem [{\citenamefont {Anwar}\ and\ \citenamefont {Ghosh}(2023)}]{anwar2023synchronization}%
  \BibitemOpen
  \bibfield  {author} {\bibinfo {author} {\bibfnamefont {M.~S.}\ \bibnamefont {Anwar}}\ and\ \bibinfo {author} {\bibfnamefont {D.}~\bibnamefont {Ghosh}},\ }\bibfield  {title} {\enquote {\bibinfo {title} {Synchronization in temporal simplicial complexes},}\ }\href {https://doi.org/10.1137/22M1525909} {\bibfield  {journal} {\bibinfo  {journal} {SIAM J. Appl. Dyn. Syst.}\ }\textbf {\bibinfo {volume} {22}},\ \bibinfo {pages} {2054--2081} (\bibinfo {year} {2023})}\BibitemShut {NoStop}%
\bibitem [{\citenamefont {Moyal}\ \emph {et~al.}(2024)\citenamefont {Moyal}, \citenamefont {Rajwani}, \citenamefont {Dutta},\ and\ \citenamefont {Jalan}}]{moyal2024rotating}%
  \BibitemOpen
  \bibfield  {author} {\bibinfo {author} {\bibfnamefont {B.}~\bibnamefont {Moyal}}, \bibinfo {author} {\bibfnamefont {P.}~\bibnamefont {Rajwani}}, \bibinfo {author} {\bibfnamefont {S.}~\bibnamefont {Dutta}},\ and\ \bibinfo {author} {\bibfnamefont {S.}~\bibnamefont {Jalan}},\ }\bibfield  {title} {\enquote {\bibinfo {title} {Rotating clusters in phase-lagged kuramoto oscillators with higher-order interactions},}\ }\href {https://doi.org/10.1103/PhysRevE.109.034211} {\bibfield  {journal} {\bibinfo  {journal} {Phys. Rev. E}\ }\textbf {\bibinfo {volume} {109}},\ \bibinfo {pages} {034211} (\bibinfo {year} {2024})}\BibitemShut {NoStop}%
\bibitem [{\citenamefont {Ott}\ and\ \citenamefont {Antonsen}(2008)}]{ott2008low}%
  \BibitemOpen
  \bibfield  {author} {\bibinfo {author} {\bibfnamefont {E.}~\bibnamefont {Ott}}\ and\ \bibinfo {author} {\bibfnamefont {T.~M.}\ \bibnamefont {Antonsen}},\ }\bibfield  {title} {\enquote {\bibinfo {title} {Low dimensional behavior of large systems of globally coupled oscillators},}\ }\href {https://doi.org/10.1063/1.2930766} {\bibfield  {journal} {\bibinfo  {journal} {Chaos}\ }\textbf {\bibinfo {volume} {18}},\ \bibinfo {pages} {037113} (\bibinfo {year} {2008})}\BibitemShut {NoStop}%
\bibitem [{\citenamefont {Sharma}, \citenamefont {Rajwani},\ and\ \citenamefont {Jalan}(2024)}]{sharma2024synchronization}%
  \BibitemOpen
  \bibfield  {author} {\bibinfo {author} {\bibfnamefont {A.}~\bibnamefont {Sharma}}, \bibinfo {author} {\bibfnamefont {P.}~\bibnamefont {Rajwani}},\ and\ \bibinfo {author} {\bibfnamefont {S.}~\bibnamefont {Jalan}},\ }\bibfield  {title} {\enquote {\bibinfo {title} {Synchronization transitions in adaptive kuramoto--sakaguchi oscillators with higher-order interactions},}\ }\href {https://doi.org/10.1063/5.0224001} {\bibfield  {journal} {\bibinfo  {journal} {Chaos}\ }\textbf {\bibinfo {volume} {34}},\ \bibinfo {pages} {081103} (\bibinfo {year} {2024})}\BibitemShut {NoStop}%
\bibitem [{\citenamefont {Jenifer}, \citenamefont {Ghosh},\ and\ \citenamefont {Muruganandam}(2024)}]{Jenifer2024}%
  \BibitemOpen
  \bibfield  {author} {\bibinfo {author} {\bibfnamefont {S.~N.}\ \bibnamefont {Jenifer}}, \bibinfo {author} {\bibfnamefont {D.}~\bibnamefont {Ghosh}},\ and\ \bibinfo {author} {\bibfnamefont {P.}~\bibnamefont {Muruganandam}},\ }\bibfield  {title} {\enquote {\bibinfo {title} {Synchronizability in randomized weighted simplicial complexes},}\ }\href {https://doi.org/10.1103/PhysRevE.109.054302} {\bibfield  {journal} {\bibinfo  {journal} {Phys. Rev. E}\ }\textbf {\bibinfo {volume} {109}},\ \bibinfo {pages} {054302} (\bibinfo {year} {2024})}\BibitemShut {NoStop}%
\end{thebibliography}

%

\end{document}